\documentstyle[aps,prl,epsf]{revtex}
\begin{document}

\draft

\title{Collective Absorption Dynamics and Enhancement in Deformed Targets}

\author{Hartmut Ruhl$^1$, Peter Mulser$^1$, Steffen Hain$^1$,
Fulvio Cornolti$^{2,4}$, and Andrea Macchi$^{3,4}$}
\address{$^1$Theoretische Quantenelektronik, TU-Darmstadt, 
Hochschulstra\ss{}e 4A,
64289 Darmstadt, Germany\footnote{\tt Hartmut.Ruhl@physik.th-darmstadt.de}}
\address{$^2$Dipartimento di Fisica dell'Universit\'a di Pisa, Piazza 
Torricelli 1, I-56125 Pisa, Italy
\footnote{\tt cornolti@mailbox.difi.unipi.it}} 
\address{$^3$Scuola Normale Superiore, Piazza dei Cavalieri 2, 
56100 Pisa, Italy  \footnote{\tt macchi@cibs.sns.it}} 
\address{$^4$Istituto Nazionale Fisica della Materia, 
Piazza Torricelli 1, I-56125 Pisa, Italy}

\date{\today}

\maketitle

\begin{abstract}
The interaction of intense fs laser pulses with thin foils that have
an imposed deformation is compared with thick targets that develop 
bow shocks. Both target types yield good absorption. Up to $80\%$ 
absorption is obtained for a $0.2\mu\text{m}$ thick, $15$ times 
over-dense foil at $4 \cdot 10^{18}\text{W/cm}^2$. A value of $50\%$ 
is obtained for a $4 \mu\text{m}$ thick, $2$ times over-dense thick target 
at $10^{18} \text{W/cm}^2$. For comparable extension and curvature
of the laser-plasma interfaces absorption levels in both targets become 
similar. In both absorption scales weakly with intensity and density.
Energy transport in thin foils and thick targets, however, is different. 
\end{abstract}

Absorption of super-intense laser pulses in solids is based on
collective mechanisms like the Brunel effect \cite{BrunelPRL87,GibbonPRL92}, 
anomalous skin effect \cite{MulserLIWM89,YangPHP96}, or $\bbox{j} \times 
\bbox{B}$-heating \cite{Kruer} with continuous transitions from one to the 
other and wide overlaps among them. Particle-In-Cell (PIC) and Vlasov 
simulations in one dimension (1D) have shown that absorption varies 
between $5-15\%$ at normal incidence to at most $60\%$ at about 
$75^{\circ}$ incidence for irradiances not exceeding several $10^{17} 
\text{Wcm}^{-2} \mu \text{m}^2$ \cite{RuhlPHP97}. Beyond this intensity
emission of harmonics and effects of self-generated dc magnetic fields
lead to a reduction of this maximum, as well as its shift towards
smaller angles of incidence and , eventually, to the formation of 
secondary relative maxima in the angular absorption behaviour
\cite{RuhlPHL95}. The signature of collective absorption is the 
generation of jets of fast electrons in the relativistic domain. 
Owing to $E^{-3/2}$ energy scaling of the collision frequency, 
collisional absorption becomes inefficient at irradiances $I\lambda^2 
\ge 10^{17} \text{Wcm}^{-2} \mu \text{m}^2$ \cite{RozmusPHP96,PricePRL95}.

All beam photon conversion into fast electrons 
occurs over skin lengths $l_{\text{s}} \approx c/\omega_p$ much less than 
a vacuum wavelength $\lambda$, simply because the free electron current 
induced by the laser field always tends to cancel the incident field
\cite{VshivkovPHP98}. 
If so, absorption is bounded to the thin critical layer. However, the 
question arises whether the geometry of the interaction region is a 
sensitive parameter for the degree of absorption. So far this problem 
has never been investigated systematically. The question to which degree
absorption can increase in deformed targets and whether thin plasma layers 
already lead to good absorption is an interesting problem in itself, e.g. 
for better understanding the relevant interaction physics, as well as it 
is essential for applications. Three substantial applications in which 
good absorption is highly desirable are (i) the generation of collimated 
intense jets of energetic electrons, (ii) broad-band intense X-ray sources 
in thin foils (for instance for back-lighting), and (iii) the fast igniter 
scheme for ICF \cite{TabakPHP94}. 

The problems addressed can be reduced to the following three questions:
(i) How does absorption change when target deformation is naturally
present due to bow shocks and hole boring or when it is imposed as for
corrugated targets? (ii) What is the energy current dynamics 
(efficiency into forward and lateral directions) in such targets? 
(iii) Is there a difference between thin and thick targets? To answer 
these questions we make use of 2D2P (two spatial and two momentum
components) Vlasov simulations for reasons of low noise and high 
resolution. We will show laser light absorption enhancement 
up to $80\%$ in thin deformed foils, lateral deflection of main 
electron jet streams, and the formation of a well collimated axial 
jet in thick targets. Finally, the occurrence and relevance of Weibel 
type instabilities and self-generated current filaments will be 
presented and discussed.

In our simulations all physical variables depend on the spatial 
variables $x$ and $y$. In addition, the distribution functions 
for the ions and electrons depend on two momentum coordinates 
$p_x$ and $p_y$. A $350 \times 128 \times 51 \times 51$ grid for 
the electrons and a $350 \times 128 \times 41 \times 41$ grid for 
the ions is used. Use is made of a charge conservative numerical 
scheme.

In order to better show how energy absorption and transport are related to 
the target deformation, first we consider a preformed plasma layer with
an imposed Gaussian deformation 
in lateral direction. The deformation is given by $x(y)=\delta \exp 
\left( -(y-y_{\text{0}})^2/\gamma^2 \right)$ where $x(y)$ denotes 
the longitudinal position of the peak density, $\delta$ the deformation 
depth as indicated in plot (a) of Fig. \ref{fig:density}, and $\gamma$ 
the deformation width. We take for the radial beam diameter $5\mu 
\mbox{m}$ at full-width-half-maximum and for $\gamma=3.8 \; \mu 
\text{m}$. For $\delta$ we use $0,1,2 \; \mu \text{m}$ in our 
simulations. For the thickness of the plasma layer we take $d=
0.2\mu\text{m}$. After a 
finite rise time the laser beam intensity is kept constant. The 
transverse locations $y_{\text{0}}$ for the peak of the pulse and 
the center of the deformation coincide. 

Fractional absorption of the laser energy as a function of time 
is presented in plot (b) of Fig. \ref{fig:density}. Both, deformed 
and planar target data are plotted. Absorption starts to rise at 
$t \approx 20 \text{fs}$, and tends to saturate at $t \approx 80 
\text{fs}$. The saturation values are between $40\%$ and $80\%$. 
Absorption in the deformed thin foils investigated in our simulations 
are well above those predicted for planar plasma films or thick 
targets \cite{WilksPRL92} for comparable parameters. Simulations
for $n_{\text{e}}/n_{\text{c}}=8$ and lower intensities but equal
deformations yield minor discrepancies for the absorption values.
Hence, we find that absorption depends only weakly on density and
intensity.


To address electron transport we look at the mass and energy currents in
the thin deformed plasma layers. Plots (a) - (d) of Fig. \ref{fig:elec_spec} 
show iso-contours of the electron energy currents $\epsilon_{\text{x}}$
and $\epsilon_{\text{y}}$, and the electron distribution projections 
$f_{\text{e}} (y,p_{\text{x}})$ and $f_{\text{e}}(y,p_{\text{y}})$ where
we have integrated over variables not indicated in the arguments. The
distribution functions are obtained from a different simulation with
$n_{\text{e}}/n_{\text{c}}=20$. Counter-propagating lateral energy currents 
are found in front of the foil. As time proceeds the lateral part streaming 
out of the center grows. This is observed from the distribution function
$f_{\text{e}}(y,p_{\text{y}})$ which has slow and fast electrons flowing 
in different lateral directions. Here, the energy current out of the center
already dominates. We find that the function $f_{\text{e}}(y,p_{\text{y}})$ 
does not become quasi-steady. In opposition, $f_{\text{e}} (y,p_{\text{x}})$ 
acquires a quasi-steady state. In that sense absorption in thin plasma layers 
is due to lateral energy flow. 


Plots (a), (b), and (c) of Fig. \ref{fig:currents} show the quasi-steady
electron mass currents and magnetic field. The mass current flows into the 
center and is balanced by a lateral return current (see plot (b)). Small 
scale filaments are present in the center of the foil. The magnetic field 
filaments saturate at $16 \text{MG}$. Combining plots (a) and (b) we observe 
that the gyro-radius of the electron current after saturation is close to 
the local classical skin length $l_{\text{s}}=c/\omega_{\text{p}} \approx 
0.08 \mu \text{m}$. The structure and scale-lengths of the magnetic field 
and current patterns are consistent with those generated by a nonlinear 
Weibel instability \cite{CalifanoPRE97} driven by the magnetic repulsion 
of counter-propagating charged electron beams. As is seen from plot (a)
of Fig. \ref{fig:elec_spec} the fast electron population does not contribute
to the growth of the filamentations since the longitudinal energy current
is always positive (hence no mutual repulsion but collimation of fast 
electrons).


We now look at simulations for thick, weakly over-dense plasma foils for 
which we let the deformation be generated by the radiation pressure of 
the short pulse itself. We take $5\mu \mbox{m}$ at full-width-half-maximum 
for the radial beam diameter. For the thickness of the plasma layer we take 
$d=4 \mu\text{m}$ while the density is $n_e/n_c=2$. After a finite rise 
time the laser beam intensity is kept constant. Plots (a) - (c) of Fig. 
\ref{fig:mag_channel} show results from a simulation of a thick foil of 
hydrogen ions. A collisionless quasi-steady bow shock is generated 
\cite{PukhovPRL97} with a shock speed of approximately $10^7 \text{m/s}$. 
The high shock speed generates a quasi-steady magnetic wake behind the 
shock-vacuum interface. The ion shock yields a deformed laser-plasma 
interface with $\delta \approx 0.5 \mu \text{m}$. Fractional absorption 
is about $50\%$ which agrees with the values obtained from the foil 
simulations with comparable deformation. We note that the density at the 
shock front is now much lower ($\approx 3 n_c$). At the shock interface 
fast electrons in longitudinal and lateral directions are generated as 
in plots (a) and (b) of Fig. \ref{fig:elec_spec} and schematically indicated 
in Fig. \ref{fig:deposition}. The energy currents $\epsilon_x$ and 
$\epsilon_y$ are of similar magnitude. In opposition to the thin foil 
case, however, fast lateral electrons are now captured by the magnetic 
field and penetrate deeply into the plasma in front of the shock. There,
they generate a magnetic field \cite{PukhovPRL97} which further collimates
electrons (see Fig.\ref{fig:mag_channel}(b)). Looking at the energy 
currents in the plasma we observe that almost the total deposited laser 
flux is converted into fast longitudinal electrons that propagate in the
channel. For reasons of quasi-neutrality a return mass current is drawn 
(see Fig. \ref{fig:mag_channel}(c)).


Instabilities may prevent the evolution of a symmetric magnetic channel
along pulse propagation direction due to vacuum-plasma interface 
distortion and bending of the magnetic channel. However, high shock 
speeds and small laser diameters help to avoid the growth of 
Rayleigh-Taylor-like instabilities due to fast mass replacement by 
fast lateral ablation (ablative stabilisation). Filamentation instabilities 
are expected to be more relevant (see plots (a) and (b) of Fig. 
\ref{fig:currents} for the case of thin foils). For counter-propagating, 
overlapping currents of similar magnitude they grow close to 
$\omega_{\text{p}}$ \cite{CalifanoPRE97}. However, the energy current 
in the channel is unidirectional since it is not balanced by an energy
return current (the fast electron population is not balanced). Hence, it 
has a collimating effect on the mass current. This effect can be
verified by imposing mirror reflecting boundary conditions. Now the
energy current is balanced and we do not obtain a magnetic channel but
magnetic bubbles (filamentation) in the plasma \cite{RuhlGSI98}.


To summarize our numerical results, we have related laser deposition and 
target geometry. Thin and thick foils both yield 
a substantial increase of short pulse absorption for moderate target
deformation. For comparable lateral extension and curvature of the 
laser-plasma interfaces we have obtained similar levels of laser 
deposition ($\approx 50\%$), no matter whether the deformation has 
been natural or imposed. We have found that absorption depends weakly 
on density and intensity but strongly on the shape of the interface. 
In both, thin and thick deformed targets fast electrons propagating 
into the center of the curved laser-plasma interface are generated. 
However, in thin plasma layers fast electrons cannot escape from the 
foil and thus heat the plasma in lateral direction. For thick foils with 
comparable lateral extension and curvature of the laser-plasma interface 
the fast electrons are captured and collimated by a magnetic field that 
extends deep into the plasma and enhances the penetration depth of the 
electrons. The plasma is now heated in pulse propagation direction. 
A high fraction of the total deposited laser flux is converted into 
longitudinal electron energy current. 

We note that there are experimental indications for both transport mechanisms.
In the experiment of Feurer {et al.} \cite{FeurerPRE97} evidence for 
significant laser-induced surface modification is given; at the same time, 
absorption is high ($\approx 45\%$). Measurements also suggest the existence 
of two electron populations, one with an energy of $\approx 400 \text{keV}$ 
and another with a few tens of $\text{keV}$, but velocity mostly parallel 
to the target surface. This is similar to what is found in our simulations 
(see plot (c) and (d) in Fig.\ref{fig:elec_spec}). Comparison of our
simulations with the experimental results suggest that both, high absorption 
and transversely flowing electrons are related to the surface deformation. 

Tatarakis {\em et al.}  \cite{TatarakisPRL98} observe a collimated plasma jet 
emitted from the rear of a thick solid density target. Their interferometric
measurements show that the laser interacts with an expanding plasma with a 
longitudinal extension of some $\mu\text{m}$. Therefore the interaction 
conditions are similar to our thick target simulations. These latter support
the conclusion by Tatarakis {\em et al.} that the fast electrons are 
collimated by magnetic fields.

The present work has been sup\-ported by the Euro\-pean Com\-mission 
through the TMR network SILASI, contract No. ERBFMRX-CT96-0043.  Use of the 
Cray T3E at CINECA was supported by INFM through a CPU grant. The authors 
are grateful for the usage of the computing facilities at ILE/Osaka. In 
particular we acknowledge Y. Fukuda and M. Okamoto as well as the CINECA 
staff for their valuable technical help.

\begin{figure}
\caption[]{Plot (a): Quasi-steady electron density for $\delta=1 \mu 
\text{m}$ at  $t=0 \text{fs}$. Blue contour areas indicate low density 
and yellow ones high density. The orange solid and dashed lines give 
ensity values along $x=1.0 \mu \text{m}$ and $y=4.96 \mu\text{m}$. The 
density is normalized to the critical electron density. Plot (b): Fractional 
absorption vs. time, for $\delta=0 \mu \text{m}$ (solid), $\delta=1 \mu 
\text{m}$ (dashed), and $\delta=2 \mu \text{m}$ (dot dot dot dashed). The 
parameters are $I\lambda^2=4.0 \cdot 10^{18} \text{Wcm}^{-2} \mu \text{m}^2$, 
$n_e/n_c=15.0$, $m_{\text{i}}=8.0 \cdot 10^{-27} \mbox{kg}$.}
\label{fig:density}
\end{figure}

\begin{figure}
\caption[]{Plots (a) and (b): Cycle-averaged electron energy current 
densities $\epsilon_{\text{x}}$ and $\epsilon_{\text{y}}$. Yellow 
areas show negative and blue areas positive values. The lines 
in (a) are along $x=1.95  \mu \text{m}$ (solid) and $y=4.96 \mu\text{m}$ 
(dashed) and in (b) along $x=1.71 \mu \text{m}$ (solid) and $y=3.94 
\mu\text{m}$ (dashed). The density is $n_e/n_c=15.0$. Plots (c) and 
(d): Cycle-averaged $f_{\text{e}}(y,p_{\text{x}})$ and $f_{\text{e}}
(y,p_{\text{y}})$. The center of the foil is at $y=4 \mu \text{m}$. 
The total lateral width for this simulation is $8 \mu \text{m}$. The 
density is $n_e/n_c=20.0$. The parameters are $I\lambda^2=4.0 \cdot 
10^{18} \text{Wcm}^{-2} \mu \text{m}^2$, $m_{\text{i}}=8.0 \cdot 
10^{-27} \mbox{kg}$, $\delta=1\mu \text{m}$, and $t=66 \text{fs}$.}
\label{fig:elec_spec}
\end{figure}

\begin{figure}
\caption[]{Quasi-steady longitudinal mass current density $j_{\text{xe}}$ 
(a), transverse mass current density $j_{\text{ye}}$ (b), and magnetic field 
(c). White contour areas are positive and black areas negative. The lines in
(a) are along $x=1.95  \mu \text{m}$ (solid) and $y=3.94 \mu\text{m}$ 
(dashed), in (b) along $x=2.0 \mu \text{m}$ (solid) and $y=3.94 \mu\text{m}$ 
(dashed), and in (c) along $x=2.0  \mu \text{m}$ (solid) and $y=3.94 \mu
\text{m}$ (dashed). The parameters are $I\lambda^2=4.0 \cdot 10^{18} 
\text{Wcm}^{-2} \mu \text{m}^2$, $n_e/n_c=15.0$, $m_{\text{i}}=8.0 \cdot 
10^{-27} \mbox{kg}$, $t=66 \text{fs}$, $\delta=1\mu \text{m}$, $j_{\text{0}}
=9.15 \cdot 10^{15} \text{A/m}^{-2}$ and $B_{\text{0}}=1.59 \cdot 10^{3} 
\text{Vs/m}^2$.}
\label{fig:currents}
\end{figure}

\begin{figure}
\caption[]{Laser irradiated thick foil. Plot (a) is the ion density, (b)
the quasi-steady magnetic field, and (c) the quasi-steady current density 
$j_{\text{xe}}$. Yellow contour areas are positive and blue areas 
negative. The lines in (a) are along $x=1.83  \mu \text{m}$ (solid) 
and $y=4.97 \mu\text{m}$ (dashed), in (b) along $x=3.5 \mu \text{m}$ 
(solid) and $y=3.94 \mu\text{m}$ (dashed), and in (c) along $x=3.34  
\mu \text{m}$ (solid) and $y=4.97 \mu \text{m}$ (dashed). The white
rectangle in (b) is schematically magnified in plots (a) and (b) of Fig.
\ref{fig:deposition}. The parameters are $I\lambda^2=10^{18} 
\text{Wcm}^{-2} \mu \text{m}^2$, $n_e/n_c=2.0$, $m_{\text{i}}=
10^{-27} \mbox{kg}$, $t=110 \text{fs}$, $j_{\text{0}}=4.58 \cdot 10^{14} 
\text{A/m}^{-2}$ and $B_{\text{0}}=2.92 \cdot 10^{2} \text{Vs/m}^2$.}
\label{fig:mag_channel}
\end{figure}

\begin{figure}
\caption[]{Electron flow (a) and energy flow (b) for a thick foil.
The plots illustrate the mass and energy flow conditions present in 
the white rectangle indicated in plot (b) of Fig. \ref{fig:mag_channel}.
Yellow areas belong to positive and blue areas to negative magnetic
fields. The magnetic fields are generated by the mass current. Its
flow directions are indicated by the black arrows in (a). The energy
current is collimated by the magnetic channel and streams in forward 
direction.}
\label{fig:deposition}
\end{figure}

\end{document}